\documentclass[conference,10pt]{IEEEtran}
\usepackage{cite}
\usepackage{amsmath,amssymb,amsfonts}
\usepackage{algorithmic}
\usepackage{graphicx}
\usepackage{textcomp}
\usepackage{xcolor}

\usepackage{booktabs} % For formal tables
\usepackage{algorithm}
\usepackage{subfigure}
\usepackage{graphicx}
\usepackage{comment}

\usepackage{tikz}
\usepackage{textcomp}
\usepackage{hyperref}
\usepackage{lipsum}

\newcommand\ChapterPrecis[2]{%
\begin{tikzpicture}[remember picture,overlay]
\node[anchor=north, draw=black, fill=yellow!20, inner sep=3pt, rounded corners, align=left, yshift=-#1] at (current page.north) 
{\parbox[t][1.3cm][l]{\textwidth}{\small #2}};
\end{tikzpicture}%
}

\def\BibTeX{{\rm B\kern-.05em{\sc i\kern-.025em b}\kern-.08em
    T\kern-.1667em\lower.7ex\hbox{E}\kern-.125emX}}

\begin{document}

\title{Pump and Dumps in the Bitcoin Era: Real Time Detection of Cryptocurrency Market Manipulations
}

\begin{comment}
\author{\IEEEauthorblockN{1\textsuperscript{st} Given Name Surname}
\IEEEauthorblockA{\textit{dept. name of organization (of Aff.)} \\
\textit{name of organization (of Aff.)}\\
City, Country \\
email address or ORCID}
\and
\IEEEauthorblockN{2\textsuperscript{nd} Given Name Surname}
\IEEEauthorblockA{\textit{dept. name of organization (of Aff.)} \\
\textit{name of organization (of Aff.)}\\
City, Country \\
email address or ORCID}
\and
\IEEEauthorblockN{3\textsuperscript{rd} Given Name Surname}
\IEEEauthorblockA{\textit{dept. name of organization (of Aff.)} \\
\textit{name of organization (of Aff.)}\\
City, Country \\
email address or ORCID}
\and
\IEEEauthorblockN{4\textsuperscript{th} Given Name Surname}
\IEEEauthorblockA{\textit{dept. name of organization (of Aff.)} \\
\textit{name of organization (of Aff.)}\\
City, Country \\
email address or ORCID}
\and
\IEEEauthorblockN{5\textsuperscript{th} Given Name Surname}
\IEEEauthorblockA{\textit{dept. name of organization (of Aff.)} \\
\textit{name of organization (of Aff.)}\\
City, Country \\
email address or ORCID}
\and
\IEEEauthorblockN{6\textsuperscript{th} Given Name Surname}
\IEEEauthorblockA{\textit{dept. name of organization (of Aff.)} \\
\textit{name of organization (of Aff.)}\\
City, Country \\
email address or ORCID}
}
\end{comment}

\author{\IEEEauthorblockN{Massimo La Morgia, Alessandro Mei, Francesco Sassi, and Julinda Stefa}
\IEEEauthorblockA{Department of Computer Science, Sapienza University of Rome, Italy\\\\
Email: \{lamorgia, mei, sassi, stefa\}@di.uniroma1.it}}

\maketitle
%\copyrightnotice

%%%%% Arxiv Reference
    \ChapterPrecis{0.15cm}{If you cite this paper, please use the International Conference on Computer Communications and Networks (ICCCN) reference: \newline Massimo La Morgia, Alessandro Mei, Francesco Sassi, and Julinda Stefa. 2020.
Pump and Dumps in the Bitcoin Era: Real Time Detection of Cryptocurrency
Market Manipulations. \textit{In 2020 29th International Conference on Computer
Communications and Networks (ICCCN)}. 1–9. \url{https://doi.org/10.1109/ICCCN49398.2020.9209660}}
%%%%

\begin{abstract}
In the last years, cryptocurrencies are increasingly popular. Even people who are not experts have started to invest in these securities and nowadays cryptocurrency exchanges process transactions for over 100 billion US dollars per month. However, many cryptocurrencies have low liquidity and therefore they are highly prone to market manipulation schemes. 

In this paper, we perform an in-depth analysis of pump and dump schemes organized by communities over the Internet. We observe how these communities are organized and how they carry out the fraud. 
Then, we report on two case studies related to pump and dump groups. Lastly, we introduce an approach to detect the fraud in real time that outperforms the current state of the art, so to help investors stay out of the market when a pump and dump scheme is in action.

\end{abstract}


\section{Introduction}

Pump and dump is a market manipulation fraud that consists in artificially inflating the price of an owned security and then selling it at a much higher price to other investors~\cite{kyle2008define,kramer2005way}. This fraud is as old as the stock market. One of the most famous pump and dumps of Wall Street history happened in the late '20. The security was the RCA Corporation, the manufacturer of the first all electric phonograph. At that time, one of the hottest pieces of technology. The fraud was organized by the "Radio Pool", a group of investors that artificially pumped RCA to the incredible price of \$549, and then dumped the shares making the price plummet to under \$10. A large number of investors lost all of their savings in this operation.
Communication was done through the radio, tabloids, and word of mouth. In the Bitcoin era, pump and dumps are more vital than ever. Indeed, communication is done through the Internet and the Web, and the targets are the hectic and almost non-regulated markets of cryptocurrencies.

The most common way to buy cryptocurrencies is through a cryptocurrency exchange. Exchanges convert fiat currencies into cryptocurrencies, and cryptocurrencies between themselves.

A cryptocurrency exchange works exactly like the traditional stock exchange. There are now hundreds of cryptocurrencies, the market is not strictly regulated, and prices are easy to manipulate. So, pump and dumps on these securities are incredibly common, with public groups in the Internet, rules, and precise and complex organization. One of the first cases known to the public involves cryptocurrency prophet John McAfee, who is one of the defendant in a complicated case of alleged pump and dump~\cite{McAfeePump}. The suit is linked to an investigation of the US SEC, a US financial regulator. Now, pump and dumps are led by a large number of self-organized groups over the Internet, and the phenomenon is viral though still not very well known.

In this work we describe the pump and dump phenomenon in the cryptocurrency ecosystem. 
We present two relevant case studies. In the first, we perform a longitudinal analysis of the gathered pump and dumps on $4$ different exchanges. In the second, we focus on Big Pump Signal, the biggest group we found in our research. Big Pump Signal is a group that works on Binance, able to generate a volume of transactions of $5,176$ BTC in a single operation, higher than the $534$ BTC volume generated together by all the pump and dumps scheme arranged on Cryptopia, YoBit and Bittrex according  to~\cite{xu2018anatomy}.
Lastly, we introduce a novel detection algorithm that works in real-time.
The algorithm is not just based on the detection of the abrupt rise of the price. The fundamental idea is to leverage the abnormal growth of so-called \emph{market buy orders}, buy orders that are used when the investor wants to buy extremely quickly, whatever the price is. Just like the colluding members of a pump and dump group when the pump starts. We show that our real-time detector outperforms the current state of the art~\cite{kamps2018moon} in a significant way, improving the expected speed of the detection from 30 minutes to 25 seconds and, at the same time, the F1-score from 60.5\% to 92\%.

\section{Pump and dump groups}
Pump and dumps are performed by self-organized groups of people over the Internet. These groups arrange the frauds out in the open on the Telegram~\cite{Telegram} instant messaging platform or Discord server~\cite{Discord}, thus everyone can join the groups without prior authorization. 
During our longitudinal research, from July $2017$ to January $2019$, we joined and daily followed all the activities performed by more than $100$ groups. Being member of the groups allowed us to retrieve and collect one of a kind information such as internal group organization, the phases of pump and dump arrangement and how the groups attract outside investors inside the market. Table~\ref{tab:group_users} shows some metrics and characteristics of $8$ representative groups we joined. 
In the following section, we report on the findings we discovered about these communities.

\begin{table*}
	\centering
	\small
    \caption{%
        Metrics of pump and dump groups
    }\label{tab:group_users}
    \begin{tabular}{l r r r r r r r}
        \toprule
        Group name &  Telegram Users &Discord Users & Hierarchy & Main Exchange & PnD (\#) & avg. Volume (\$)\\
        \midrule
        Big Pump Signal &                      $72,097$ & 104,830 & affiliation &  Binance & 32 & 7,245,437 \\

        Trading Crypto Guide &              $91,725$ & --- & vip &  Binance  &  17 & 2,442,923 \\

	  Crypto Coin B &                         $166,689$ & --- & vip &  Binance & 6 & 5,733,637 \\

        Crypto4Pumps &                        $11,716$ & --- & vip &  Bittrex  & 47 & 491,395 \\

        Pump King Community &            $7,771$   & --- & vip &  Bittrex  &  18 & 931,960 \\

        Crypto Family Pumps &               $4,449$   & 5,299 & free &  Cryptopia   & 28 & 23,800 \\

        Luxurious pumps &                     $6,020$   & --- & free &  YoBit  & 16 & 4,997 \\     

        AltTheWay &                              $7,333$   & --- & free &  YoBit  & 89 & 700 \\    
        \bottomrule
    \end{tabular}
    
\end{table*}

\subsection{Group organization}
\begin{figure}
\includegraphics[width=0.48\textwidth]{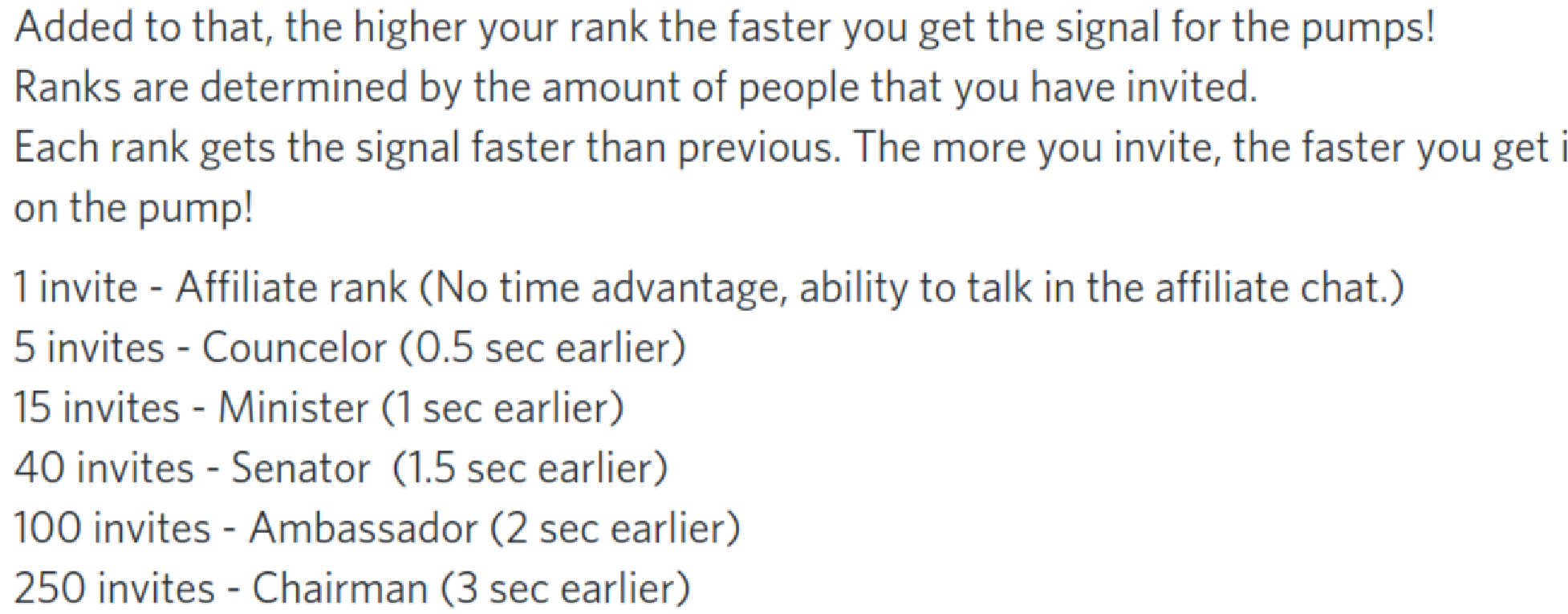}
\caption{Screenshot of a Discord Info room, where it is explained how the affiliation system works.}
\label{fig:discord}
\end{figure}

Pump and dumps groups have leaders (or admins) that administrate the group, and a hierarchy of members. If a member is higher in the hierarchy, he gets the message that starts the pump by revealing the target cryptocurrency a few moments earlier than lower ranked members. This way, the member has higher probability to buy at a lower price and make more money out of the pump and dump operation. The advantage in terms of time of being at a higher level is usually between~$0.5$ and~$1$ second with respect to the next level, and the maximum advantage is in the interval between~$3$ and~$8$ seconds. Most groups are organized as an affiliation system ---climbing the hierarchy is possible by bringing new people to the group. The larger is the number of new members brought to the group, the higher the ranking. Fig.~\ref{fig:discord} shows the affiliation system of Big Pump Signal group and the rank's benefits.

Some groups have a simpler hierarchy with only two levels: Common members and VIP members. In these groups, to become a VIP the user has to pay a fee, usually in Bitcoins, in the range of $0.01$ to $0.1$ Bitcoins (from approximately 71 to 710 USD at current exchange rates\footnote{Data retrieved on January 10, 2020}). 
In the pump and dump groups, the admins are the only people that take decisions. We saw only in rare cases the admins running polls to agree on the hour of the pump or the exchange to use, and never to decide the target cryptocurrency. 

\subsection{Group communication}

To communicate and organize the pump, the groups typically use Discord servers and Telegram channels. Telegram is an instant messaging service, and a Telegram channel is a special kind of a chat in which only the owner of the channel can broadcast public messages to all the members. Discord is a VoIP and text chat service. It was originally designed for video gaming communities, but nowadays it is widely used by other communities as well. 

Discord offers the possibility to create macro sections and host multiple chat rooms. Each section has its own topic. In our analysis we have found that all the pump and dump Discord servers are organized in roughly the same fashion, with the following sections:
\begin{itemize}
  \item \textbf{Info \& How-Tos}: These two sections are like an electronic bulletin board with pinned messages. Both sections are composed of several rooms that contain only one or very few messages. The rooms of the Info section usually contain the rules of the group, the news about the group, how the affiliation system works (Fig.~\ref{fig:discord}), and the F.A.Q.. The rooms of the How-Tos section contain manuals related to the cryptocurrency world or the best practices to participate in a pump and dump operation.
  \item \textbf{Invite}: This section contains rooms where the bots of the server live. Here, the users can query the bots in order to generate invite links to bring new members or to know the number of people that joined the server by using their invite links.
  \item \textbf{Signal}: This is the core section of the group, in which only the admins can write. Inside this section there are usually two rooms: The pump-signal, and the trading-signal. In the first room, the admins share info about the next pump and dump operation. In the second, they share trading advices.  
  \item \textbf{Discussion}: in this section, there are rooms covering different topics where the group members can freely chat.
\end{itemize}
Usually, the messages written in the news and in the pump-signal rooms are broadcasted to the Telegram channel as well.

\subsection{Organization of the pump and dump operations}

\begin{figure}
\includegraphics[width=0.49\textwidth]{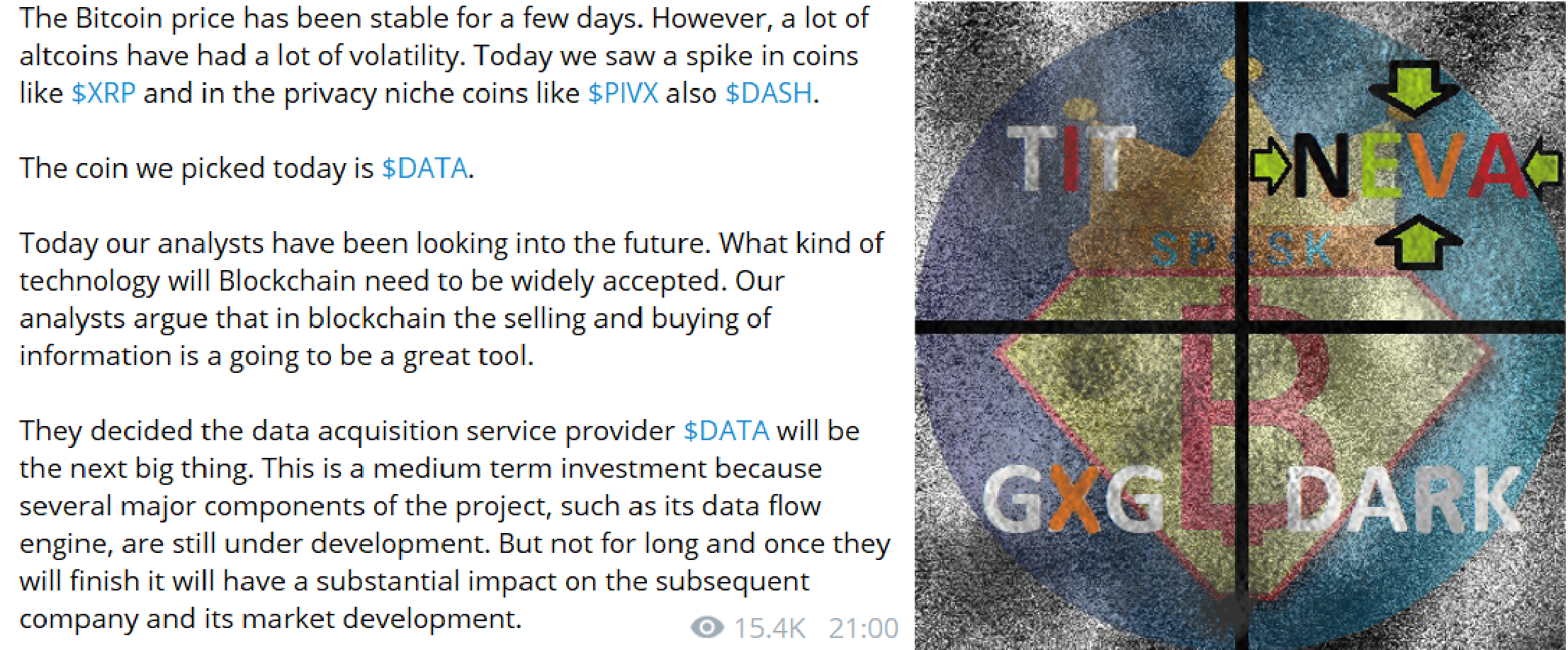}
\caption{Messages that indicate the start of a pump and dump operation on the Streamr DATAcoin (on the left) and the NevaCoin (on the right).}
\label{fig:coin_signal}
\end{figure}

The levels of activity of the many pump and dump groups in the Internet differ considerably. The most active ones perform roughly one pump and dump operation a day. Less active groups perform one operation a week. Other groups perform operations only when they believe the market conditions are good. The steps during the operation are typically as follows:
\begin{itemize}
\item A few days or hours before the operation the admins announce that the pump and dump will happen and communicate which is the exchange that will be used, the exact starting time of the operation, and whether the operation will be FFA (Free for All---everybody gets the message at the same time) or Ranked (VIPs and members of higher levels in the hierarchy get the starting message before the other members).
\item The announce is repeated several times, more frequently as the starting time of the operation gets closer.

\item When the pump starts, the target cryptocurrency is revealed to the members of the group. The exact time depends on the position in the hierarchy. Usually, the name of the cryptocurrency is contained in an image that is obfuscated in a way that only humans can read it quickly. Fig.~\ref{fig:coin_signal} shows an example, a message that instructs to start a pump and dump operation on the NevaCoin. The idea behind the obfuscation is to make it hard for bots to parse the message with OCR techniques and start the market operations faster than humans. 
\item Lastly, a few seconds after the start of the pump, the admins share a tweet or a news and invite all the members of the group to spread the information that the price of the cryptocurrency is rising. This is done in dedicated chat boxes, forums, and Twitter. The goal of this activity is to create so-called FOMO (Fear of Missing Out) of a good opportunity of investment and attract investors from outside the group.

\end{itemize}

\section{Case study}
\subsection{The groups, the pump and dumps, and the exchanges}
In this section, we describe an in-depth investigation on the cryptocurrencies and the exchanges used for the pump and dumps. We do so in a period of time that goes from July $2017$ to January $2019$. In this period, we found more than $100$ groups, by keyword search (e.g.: \textit{"Pump","Dump","Signal"}) on Telegram, Twitter, Reddit, BitcoinTalk~\cite{bitcointalk}, or manually extracting information from CoinDetect~\cite{coindetect} or the PADL~\cite{padl} Android app.
From this set, we select $19$ different groups, since the other are no longer active, only broadcast event from other groups, or the number of users was quite small. Reading the Telegram channel history of these groups, we found $343$ pump and dump operations carried out on $4$ exchanges. For all the pump and dumps collected, we retrieved the historical trading data, as much detailed as we can, scraping the exchanges' website or via APIs.
Analyzing our data, we found that the scheme involved $194$ different cryptocurrencies, only $143$ of which are still listed by the Coinmarketcap site.
It is quite common that coins disappear from the cryptocurrency world. Indeed, by analyzing the volumes in the last $24$ hours of the cryptocurrencies that are still active, we found out that $112$ of them moved less than $\$1$ million in total in all the exchanges in which they are listed. Actually, $46$ of them moved less than $\$10,000$.

Also, $100$ of the coins used for pump and dumps are below $20$ million dollars of market capitalization, with $34$ of them being below $1$ million. The first asset with less than $20$ million dollars is at the $220th$ position of the cryptocurrency ranking by market capitalization. So, the targets of pump and dumps have a very low net worth value and a huge amount of circulating supply. Lastly, we find that $99$ cryptocurrencies out of $141$ are priced below $0.4$ dollars. As such, with a relatively small investment, pump and dump groups can buy huge amounts of shares and easily increase their price in the pump phase of the fraud.

YoBit is the exchange where most of pump and dump operations happen, while Binance is the most popular exchange among all groups. 
Each pump and dump group tends to use the same exchange. Indeed, if the groups jumped from one exchange to the other, the members would be forced to move their assets accordingly and pay the related fees. 
The pump and dumps on currencies with higher market capitalization are typically carried out on Binance, the ones with lower market capitalization on Cryptopia.
In particular, the median market capitalization of the cryptocurrencies for exchange is $\$25,574,192$ for Binance, $\$2,619,703$ for YoBit, $\$2,512,627$ for BitTrex, and $\$144,373$ for Cryptopia.

Finally, we notice that pump and dump schemes do not affect only the target exchange. Indeed, the cryptocurrency markets, like the real stock exchange markets, are constantly under monitoring of arbitrage bots that look for profitable trading. Arbitrage is a practice that consists of taking advantage of the price difference between two markets by buying in one market and selling in another market at a higher price. Some seconds after a pump and dump starts, the price of the coin under attack quickly increases its price. The price spike triggers the arbitrage bots that start to buy and sell the currency on a different market or other trading pairs. Table~\ref{tab:arbitrage} reports the values during the pump and dump carried out on the Streamr DATAcoin (DATA) of the 29th of September 2018 for different exchanges. As we can see, the biggest transaction volumes and the highest prices are reached on Binance. Indeed, it was the exchange used for the operation. However, also the other exchanges record a rise of the price in the same time range. 
\begin{table}
\small
\centering
    \caption{%
        Effect of arbitrage.
    }\label{tab:arbitrage}
    \begin{tabular}{l c r r r}
        \toprule
        Trading pair & Exchange &   Volume (\$) & Open (\$)& Max(\$)\\
        \midrule
        DATA/BTC &              Binance & $5,732,909$ & $0.0354$ & $0.0916$ \\

        DATA/ETH &              Binance & $568,722$ & $0.0353$ & $0.0924$  \\

        DATA/BTC &              Bitfinex & $103,318$ & $0.0353 $& $0.0564$ \\

        DATA/ETH &              Bitfinex & $62,640$ & $0.0333$ & $0.0552$ \\

        DATA/USD &              Bitfinex & $175,274$ & $0.0349$ & $0.0551$ \\

        DATA/BTC &              HitBTC & $27,502$ & $0.0359$ & $0.0564$ \\

        DATA/ETH &              HitBTC & $5,084$ & $0.0330$ & $0.0735$ \\

        DATA/USDT &            Gateio & $31,357$ & $0.0347$ & $0.0600$ \\

        \bottomrule
    \end{tabular}
\end{table}

\subsection{The Big Pump Signal group}
\begin{figure}

 \includegraphics[width=0.45\textwidth]{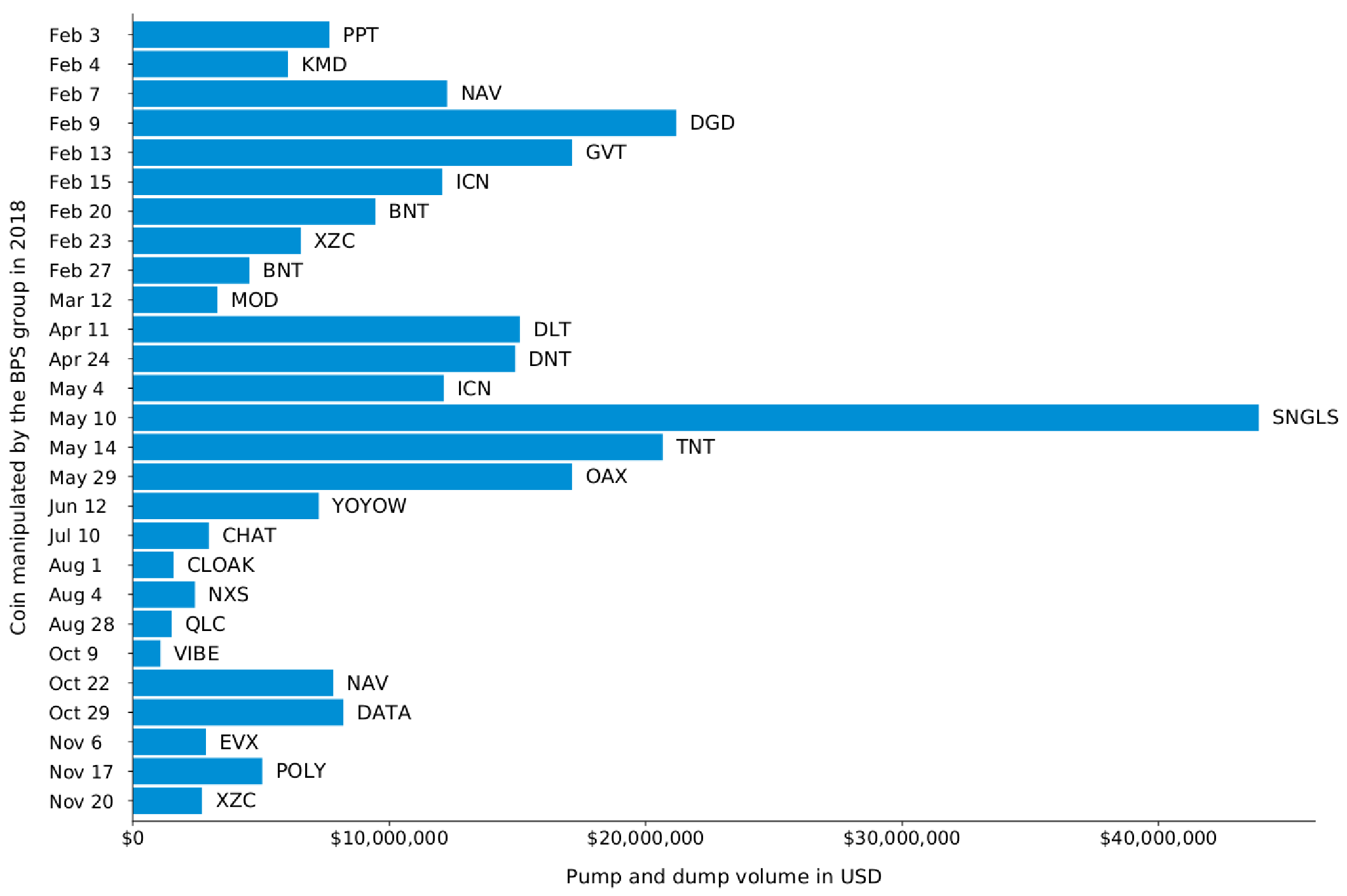}
  \caption{Big Pump Signal pump and dumps}
  \label{fig:BPS_pumps}
  \end{figure}

With more than $104,000$ members on Discord and more than $72,000$\footnote{Data retrieved on January 2019} members on Telegram, Big Pump Signal (BPS) started in December 2017 over Telegram and is arguably the largest pump and dump public community in the Internet. Reading the pump announcements on the Telegram channel of Big Pump Signal, we found 32 pumps, 27 of which carried out on Binance (see Figure~\ref{fig:BPS_pumps}) and 5 on Cryptopia. 
Throughout all their pump and dump operations, the group moved $82,369,542$ USD globally, if we count the first $6$ minutes of every pump, and $267,482,773$ USD globally, if we count the whole coin oscillation due to the pump. Their most successfully pump and dump was organized on May 10, 2018, when they targeted the SingularDTV (SNGLS) alt-coin. In this operation, the value of the SNGLS coins sharply oscillated for more than $9$ hours and recorded a volume of trades of $5,176$ Bitcoin ($36,750,000$ USDs).
BPS has an affiliation hierarchy, the highest level of which is achievable after inviting $250$ new members. In ranked pump and dump operations, the affiliation guarantees the members to receive the signal $3$ seconds before the unranked members. The Big Pump Signalers have, since the beginning, promoted the group in many ways: Advertisement on main social networks like Twitter and Quora, through the affiliation systems, and, by organizing lotteries inside the groups with prizes in Bitcoin or Ethereum. Thanks to the aggressive marketing campaigns and the hype on the cryptocurrencies in late $2017$, the Big Pump Signal group has grown extremely fast reaching around $200,000$ members in January $2018$.

\subsubsection{Organization of the BPS operations}

The organization of the BPS pump and dumps steadily improved with time. In the beginning, they operated on a single coin, while double trading on the Bitcoin and Ethereum target pairs. The target coin was revealed to the members in plain text. They were also given a target price with the goal of pushing the pump up to the target so to attract external investors and, afterwards, profit by selling to them. However, this strategy was the opposite of being optimal for the admins: They noticed that greedy group members started to sell below the target price value, forbidding the pump and dump operation to reach the highest peaks. So, they stopped indicating the target price and carried the purchasing power in a single trading pair to get maximum gain. 

Successively, the admins discovered the presence of bots inside the channel. The bots were very fast at buying, much faster than the group members. To mitigate their effect, the admins changed the pump and dump signal--from a text, to an obfuscated image--so that it was extremely difficult for bots to read.

As the group grew bigger, the admins started targeting also cryptocurrencies with medium market capitalization. The admins claim that the choice of the coin is typically based on technical analysis. They also claim to re-pump the cryptocurrencies by collaborating with a small investment firm. The investment firm is thought to be frequently involved in or to organize pump and dumps on her own. An example are the pump and dumps of the Monetha coin (MTH) and the WePower (WPR) coin on the Binance platform on September $17$, $2018$. 
Our analysis shows that BPS typically chooses cryptocurrencies that have had a steady price and news coverage in the recent past. They exploit the news coverage to generate interest and attract external investors. An example are the retweets of news from the fake twitter account of John McAfee (@oficiallmcafee, and the still reachable @TheJohnMcafee) belonging to the admins of the group. 
\begin{figure}

  \includegraphics[width=0.48\textwidth]{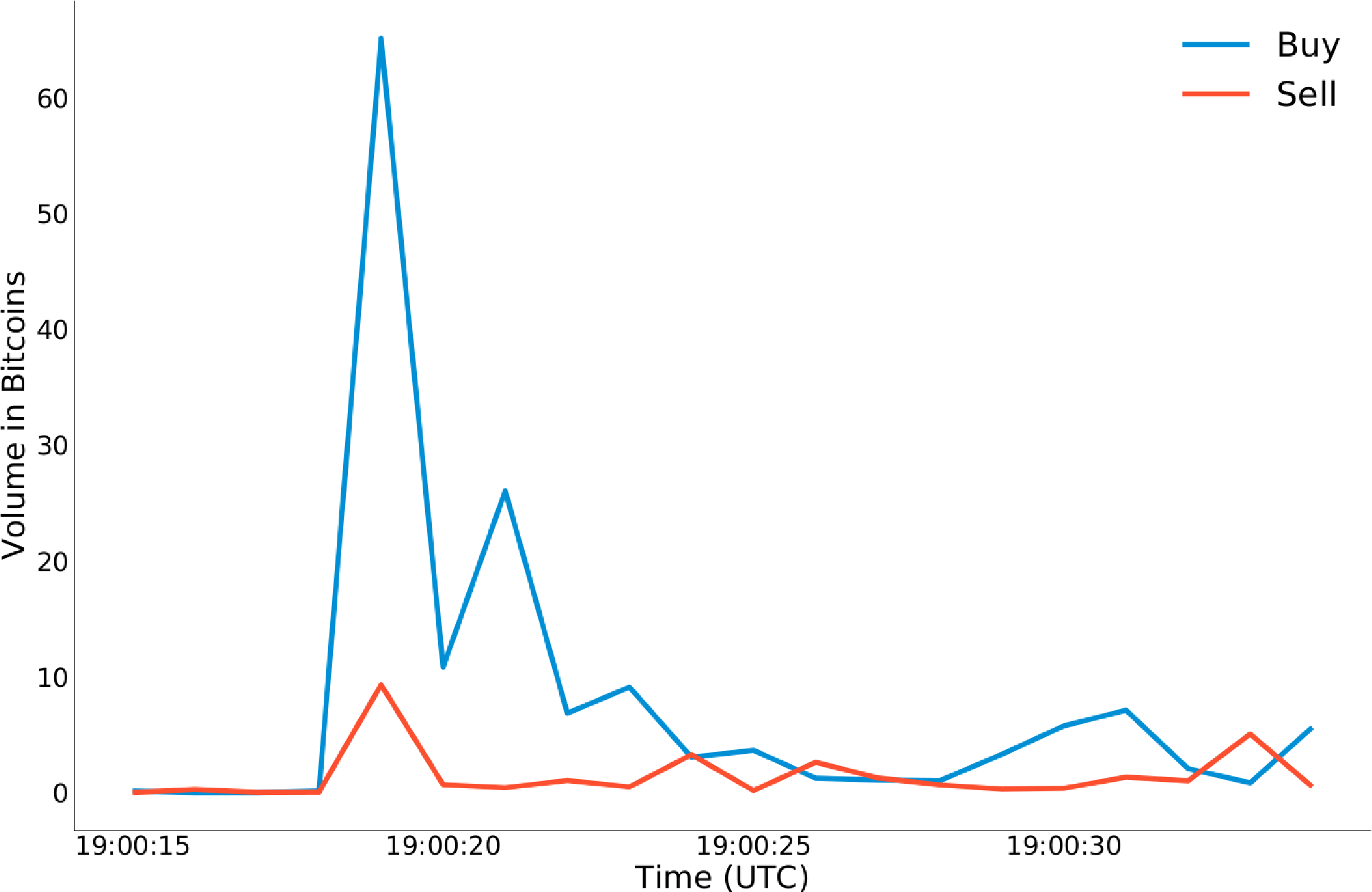}
  \caption{Pump on the OAX}
  \label{fig:BPS_oax}
  \end{figure}

\subsubsection{Analysis of the pump phase}
The BPS group moves large volumes of Bitcoins in each operation. We analyse one of these operations to investigate in depth the pump phase, the most interesting step of a pump and dump operation. Figure.~\ref{fig:BPS_oax} depicts a second by second zoomed image on the very first $30$ seconds of the pump on the OAX cryptocurrency. The upper (blue line) in the figure represents the buy volume, while the (lower) orange one the sell volume. We observe that the volume of the buys and sells in the first seconds is very close to zero. Then, there are two buy peaks (blue line in Figure~\ref{fig:BPS_oax}) of approximately $65$ Bitcoins (sec. $19$) and $26$ Bitcoins (sec. $21$) respectively. The two peaks correspond to the actions of VIPs and the common members---a normal behavior, considering that the group has a ranked policy. We also observe a peak on the sell volumes (orange line in Figure~\ref{fig:BPS_oax}) of almost $10$ Bitcoins on the moment of the first buy peak, the $19th$ second. Considering that group members are still buying and that the reaction time for outsiders is too short, this sudden big sell volume is abnormal. In fact, there can be only two possible actors to sell their shares: the bots and the admins. In order to discern among the two we need to investigate the single transactions. Our analysis shows that, as the price rises, there are many small sell operations at incremental values, probably by the arbitrage bots. Then, we observe a last single shot transaction for over $4$ Bitcoins when the OAX coin reaches the trading value of $0.00012$ Bitcoins, probably done by the admins of the group. We believe they have operated through a \textit{sell limit} trade order---a conditional order triggered when the price of a trading pair reaches/out-tops a given value. Of course, the same order could have been placed also by an outside investor. However, we believe that a sell limit for that amount and that overcomes the initial price of the $41$\% is most likely by an insider.

\section{Pump and dump detection}

\subsection{The idea}
As we know, standard investors are the victims of pump and dump schemes. When they see that the price of a cryptocurrency rise, they can believe it can be a good investment opportunity. This is not the case when a pump and dump scheme is in action---the rise does not have economical grounds, it is just market manipulation. To protect investors, it is important to understand if a pump and dump scheme can be detected, and how fast it can be done. This is the goal of this section.

To better understand how pump and dumps can be detected, it is important to have some basic notions. The pending orders for a cryptocurrency, like any other security, are listed in the \emph{order book} for that cryptocurrency. The book is a sorted double list of sell (ask) and buy (bid) orders not yet filled. The asks are sorted from the lowest price to the highest, the bids are sorted from the highest to the lowest.
%The difference between the first ask and the first bid is called \emph{spread}.
The fastest way to buy on the market is through a \emph{buy market order}. A buy market order looks up the order book and fills all the pending asks until the requested amount of currency is reached. Although a market order is completed almost instantly, the price difference between the first and the last ask needed
to fill the order can be very high, especially in markets with low liquidity, and so the price can rise considerably. A more careful investor would use
\emph{limit buy orders}, orders to buy a security at no more than a specific price. Buy market orders are not frequent in normal transactions and are typically used by investors that need a fast execution. Just like the members of pump and dump groups in action. Our idea is to use this pattern, along with other information about volume and price, to detect when a pump and dump scheme starts.

\subsection{The data}

%%AGGIUNTA 
As highlighted by Kamps et al.~\cite{kamps2018moon}, it does not exist a dataset of confirmed pump and dumps in the literature. So we need to build one for the purpose of this work. Starting from the $19$ groups we joined, we select only the pump and dump schemes carried out on Binance. We made this choice for two main reasons: The first one is that Binance exposes APIs~\cite{binanceapi} that allow to retrieve every single transaction in the whole history of a trading pair, differently from other exchanges that allow to retrieve data with a transaction granularity only for few hours back. The second is that pump and dumps on other exchanges are usually carried out by groups with few active members and economic resource, consequently they are forced to target alt-coin that have almost no volume of transactions for days before the scheme. Thus, we believe that pump and dumps carried out on Binance are the most interesting and challenging to detect.
From the initial set of pump and dumps, we select all the events that were carried out on Binance. At the end we get $104$ pump and dump events, arranged by $12$ different groups. Doing so, we retrieve the historical trading data for $7$ days before and after the event, for a total of $14$ days. Some pump and dumps were carried out a few days apart on the same alt-coin, so we discarded duplicate days. In the end, we have about $900$ days of trading.
The data are a list of trade records: Volume, price, operation type (buy or sell), and the UNIX timestamps. The trades are provided by Binance in a compressed way, this means that a record belonging to the same order at the same price have aggregated quantities, and a single order that is filled at different prices is splitted into more records.

Unfortunately, the Binance APIs do not tell the kind of order (e.g.: \textit{Market, Limit, Stop Loss}) placed by the buyer, so we need to infer this information.
To do this, we can use the fact that these kind of orders are filled in a single shot, and so we can aggregate the trades filled at the same millisecond as a single one. 
Since we do not know the original nature of these orders we define them as \textit{rush orders}. A problem of this inference method is that it misses the orders that are completely filled by the first ask of the order book. Still, we believe that we have a good witness on the abrupt rise of market orders even with this approximation.
As a contribution to the community we will publicly release this dataset.

\subsection{Features and classifiers}
\label{sec:features}
To detect the start of the fraudulent scheme, we analyze several kinds of features and use them to feed two different classifiers: Random Forest and Logistic Regression.
A Random Forest~\cite{breiman2001random} is an ensemble learning method consisting of a collection of decision tree classifiers such that each tree depends on the values of a random vector sampled independently, each tree casts a vote, and the prediction is the most popular class among all the votes. 
Logistic regression~\cite{hosmer2013applied} is a type of regression analysis used to calculate the outcome of dependent variables based on one or more independent variables estimating the parameters of a logistic model, where the value of the predictor variable is between $0$ and $1$.
We built our features upon the idea of~\cite{siris2004application} for the detection of Denial of Service attacks through an adaptive threshold. Since in our case we do not want to find a threshold, we restyle their idea in this way: We split our data in chunks of $s$ seconds, and we define a moving window of size $w$ hours.

We conduct several experiments with different sets of features and settings regarding the window and the chunk sizes. Since our goal was to build a classifier able to detect a pump and dump scheme as quickly as possible from the moment it starts, it is crucial that the chunk size is reasonably short. At the end of our study, we found that the best configuration, in terms of F1-score, was achieved with a chunk size of $25$ seconds and a window size of $7$ hours, while the fastest one with a chunk size of $5$ seconds and a windows size of $50$ minutes.
Following are reported the final features we used:
\begin{itemize}
  \item \textbf{StdRushOrders} and \textbf{AvgRushOrders}: Moving standard deviation and average of volume of rush orders in each chunk of the moving window.
  \item \textbf{StdTrades}: Moving stardard deviation of the number of trades, both buy and sell. 
  \item \textbf{StdVolumes} and \textbf{AvgVolumes}: Moving standard deviation and average of volume of trades in each chunk of the moving window.
  \item \textbf{StdPrice} and \textbf{AvgPrice}: Moving standard deviation and average of closing price.
  \item \textbf{AvgPriceMax} and \textbf{AvgPriceMin}: Moving average of maximal and minimum price in each chunk.
  \end{itemize}

Once a pump is detected we pause our classifier for $30$ minutes to avoid multiple alerts for the same event.

\subsection{The importance of rush orders}
\begin{figure} 
\includegraphics[width=0.48\textwidth]{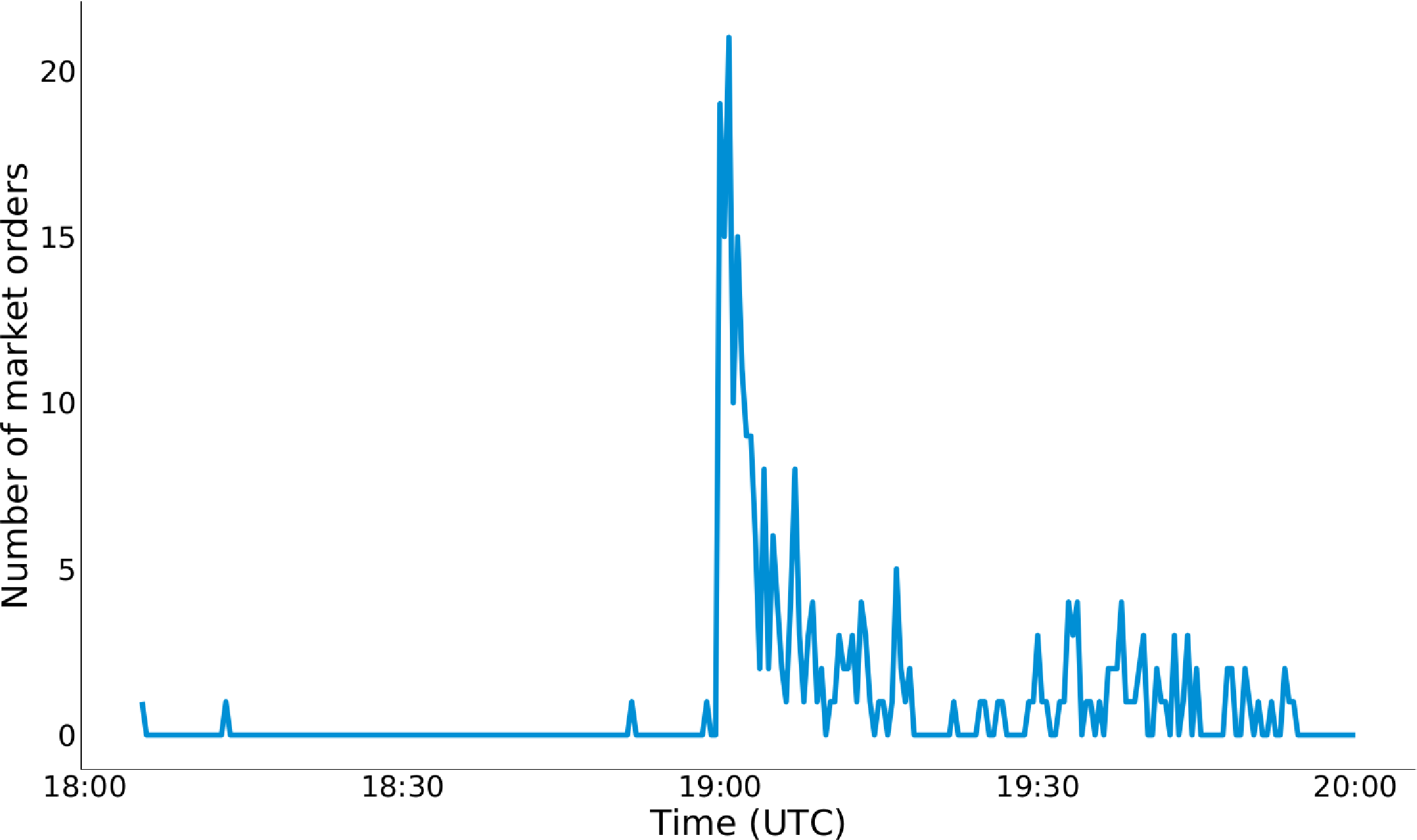}
\caption{Rush orders during a pump and dump scheme.}
\label{fig:market_op}
  \end{figure}

In this section, we explore how the rush orders are important to detect the start of a pump and dump operation.
\begin{figure} 
\includegraphics[width=0.48\textwidth]{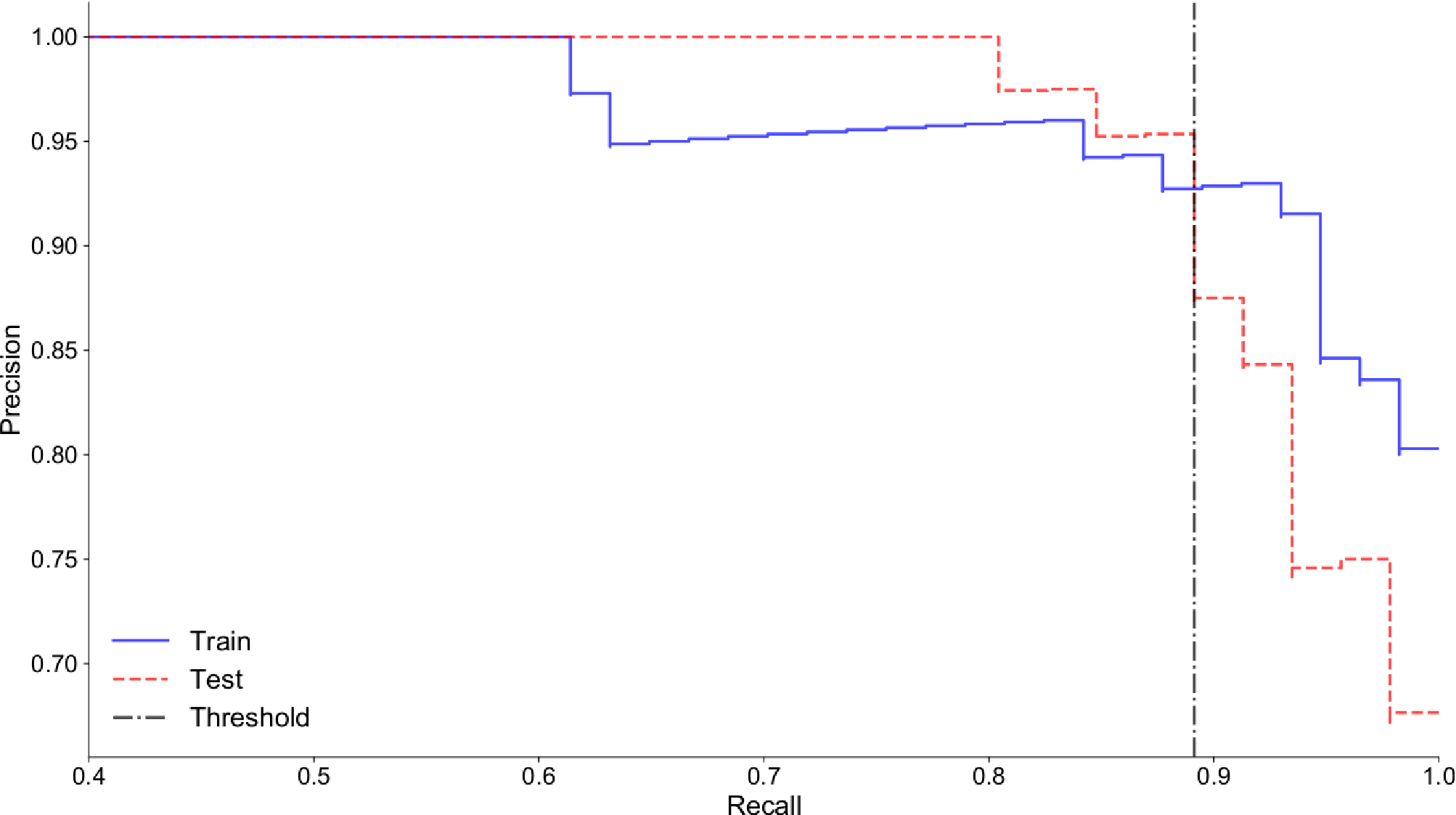}
\caption{Precision recall curve for train and test sets.}
\label{fig:pr_curve_train}
\end{figure}
Fig.~\ref{fig:market_op} shows the number of rush orders during a pump and dump scheme on the VIBE cryptocurrency on September 9th, 2018. 
As we can see, rush orders are rare during the hours before the pump and suddenly grow just at the start of the scheme. 
The goal of this experiment is to understand if the rush orders are an effective feature to detect the start of a pump and dump scheme and find a threshold beyond which classify the growth as anomalous. 
To learn the threshold, we proceed as follows: We compute the StdRushOrder feature as described in Section~\ref{sec:features}, then we label each chunk as True if the timestamp of the pump and dump signal falls into the chunk time range, False otherwise. We randomly split our dataset into the train ($50$\%) and test ($50$\%) sets, we compute the precision-recall curve for the train set, and we pick a threshold that is a tradeoff between the precision and the recall. Then we evaluate the same metrics at the picked threshold for the test set.
The result of this experiment is shown in Fig.~\ref{fig:pr_curve_train}. We choose as value for the threshold (the black dashed line in the figure) $30.32$ that provides a precision of $91.1\%$ and a recall of $89.9\%$ on the train set (the blue line). As we can see, the same threshold value provides a very similar score on the test set too (the red dashed line).
Given these good results, we can assume that the rush orders feature is an extremely good parameter to evaluate the start of a pump and dump.

\subsection{The results}

Although we retrieved $2$ weeks of data for each pump and dump scheme, initially we use only $3$ days---the day of the fraud, the day before, and the day after---since in this timeframe we can assume that no other frauds are present in the dataset. We noticed that, among the market manipulations we collected, different groups arrange schemes on the same alt-coin a few days apart. Moreover, we are aware that some groups delete the pump and dump signal from the chat history and that there do exist groups that we are not able to monitor such as groups that communicate in Chinese or Russian or in private groups.
Since our dataset consists of $104$ pump and dumps, we do not split the dataset into the standard train test sets, but we performed a $5$ and $10$ folds cross-validation, in order to get a more reliable evaluation of the performance.
\begin{table}
\small
\centering
\caption{%
Classifiers performance.
}\label{tab:classifiers}\begin{tabular}{l c r c c c}
\toprule
Classifier & Chunk size&  Precision& Recall &F1\\
\midrule
Kamps (Initial)& $1$ Hour  & $15.6$\% & $96.7$\% & $26.8$\%\\
Kamps (Balanced)& $1$ Hour  & $38.4$\% & $93.5$\% & $54.4$\%\\
Kamps (Strict)& $1$ Hour  & $50.1$\% & $75.0$\% & $60.5$\%\\
\midrule
%LR  &$25$ Sec& $5$ &  $91.5\%$  & $88.4\%$ & $89.5$\%\\
%LR  &$25$ Sec& $10$&   $93.4\%$ & $89.5\%$ & $90.8$\%\\
%\midrule
RF ($5$ Folds) &$5$ Sec  &  $92.4\%$ & $78.4\%$ & $84.0$\%\\
RF ($10$ Folds) &$5$ Sec & $92.2\%$ & $77.5\%$ & $82.7$\%\\
RF  ($5$ Folds)&$15$ Sec&  $91.3\%$ & $84.4\%$ & $87.7$\%\\
RF  ($10$ Folds)&$15$ Sec & $91.1\%$ & $83.3\%$ & $87.0$\%\\
RF  ($5$ Folds)&$25$ Sec &  $93.7\%$ & $91.3\%$ & $91.8$\%\\
RF  ($10$ Folds)&$25$ Sec & $93.1\%$ & $91.4\%$ & $92.0$\%\\
\bottomrule
\end{tabular}
\end{table}
%In Tab~\ref{tab:classifiers} are reported the average result for each experiment. 
For the Random Forest classifier we use a forest of $200$ trees, each leaf node must have at least $6$ samples, and a maximum depth of $4$ for each tree. For the Logistic Regression classifier, we used the \textit{BFGS} solver with a regularization strength of C equals to $1$.
Since we notice that the classifiers based on the Random Forest algorithm perform slightly better than the ones based on the Logistic Regression in all the chunk size, for the Logistic Regression model we report in Tab~\ref{tab:classifiers} only the results with a chunk size of 25 seconds. Moreover, from the results of the Random Forest classifier, it is possible to note the relationship between the chunk size and the performance of the classifiers. Indeed, while the precision is pretty stable in all the time frames, the recall increases as we increase the chunk size dimension.

\begin{table}
\small
\centering
\setlength{\tabcolsep}{14pt} % Default value: 6pt
\caption{%
Feature Importance.
}\label{tab:importance}\begin{tabular}{l r }

\toprule
Feature & Importance\\
\midrule
StdRushOrders  & $0.401$ \\
StdTrades  & $0.202$\\
AvgRushOrders  & $0.153$ \\
AvgVolumes  & $0.097$\\
StdVolumes   & $0.076$\\
StdPrice    & $0.026$ \\
AvgPrice    & $0.014$ \\
AvgPriceMax    & $0.014$\\
AvgPriceMin    & $0.011$\\
\bottomrule
\end{tabular}
\end{table}

In Tab~\ref{tab:importance} we list the importance, computed with the Gini Impurity, of each feature used with the Random Forest classifier. As we can see, the best features are the ones based on the rush orders and the number of trades.
Once our methodology has been finalized, we trained a 25 second detector classifier with the $3$ day dataset and used the remaining part as test looking for other suspect events. After the evaluation, we got $29$ events that we are not able to link to evidence. We believe that virtually all of them are pump and dumps, whose evidence has been deleted or organized by groups that may not be public.
\begin{figure}
\includegraphics[width=0.48\textwidth]{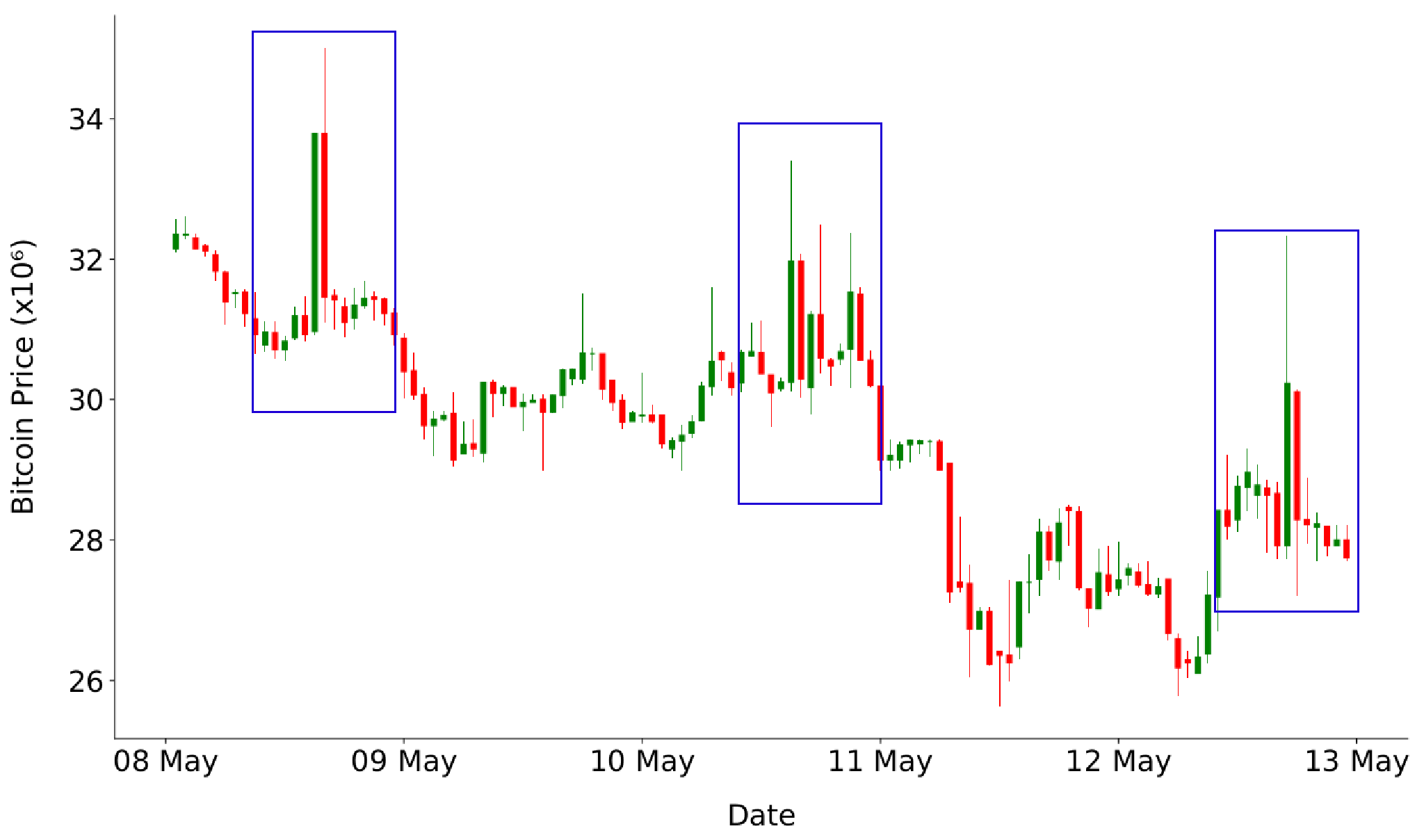}
  \caption{DLT candlestick chart}
  \label{fig:false_positive}
\end{figure}
Fig.~\ref{fig:false_positive}, for example, shows the candlestick chart for the Agrello coin (DLT), from May 8 to 13. The event in the center is a pump and dump for which we have evidence, the other two are suspects detected by the algorithm. As you can see, the behavior is almost the same, including the fact that the currency goes back to the normal price quickly (the dump). Our classifier, based on the detection of the abnormal presence of rush orders and not just on the price of the security, does a good job in detecting pump and dumps and suspect events that, anyways, the mindful investor wants to stay away from. The classifier is also very fast---in all the cases the pump and dump is detected within a timeframe of 5 seconds and, in some cases, even before the start of the pump when there is pre-pump activity in action.

\subsection{Comparison with other pump and dump detectors}
\label{subsec:comparision}
After validating our classifier, to better understand the performance of our solutions we introduce the pump and dump detector of Kamps et al.~\cite{kamps2018moon} as the baseline.
In their work, they simulate a real-time detector using as input candlesticks of 1 hour to detect pump and dumps. So, detection time can be up to 1 hour, with an expectation of 30 minutes. To detect pump and dumps their methodology exploit two anomaly thresholds one for transactions volume and the other for the coin price. They compute the values of the thresholds using average windows on the recent history of the candlestick under observation. Hence, if both the price and the volume are higher than the computed thresholds they mark the point as pump and dump event.
Finally, Kamps et al. provide 3 different parameter configurations to compute the threshold: Initial, Balanced, and Strict. The Basic configuration maximize the recall, the Strict the precision, while the Balanced is a trade off between the previous two.
In their work they only mention the number of alleged pump and dumps that their classifier detects, unfortunately they are not able to provide scores in terms of precision and recall since their dataset lacks of ground truth.

To use the Kamps et al. detector as the baseline for our task, we start replicating their classifier and testing it on their dataset, detecting the same number of the pump and dumps they declare in their work. Then, we apply their methodology on our dataset; results are shown in Table~\ref{tab:classifiers}. As we can see, all our classifiers outperform in terms of F1-score the Kamps et al. detectors. Note the difference in performance between their best configuration based on Strict parameters and our slowest classifier: Not only our performances are considerably better than theirs, we score 93.1\% of precision and 91.4\% recall against their 50.1\% precision and 75.0\% recall, but our detector is also faster. 
These results also highlight that, due to the high volatility of the cryptocurrencies market, detectors based only on the coin price and transaction volume are prone to a large number of false positives.

Differently from us, Xu et al.~\cite{xu2018anatomy} build a classifier able to predict the currency target of the next pump and dump to provide a tool for strategic trade. Since the goals are different, we can not make a comparison in terms of performance between our work and their solutions.
Indeed, they prefer to maximize the probability of gain from the investment maximizing the recall at the expense of low precision. They assume that buying wrong currencies does not affect their trading strategy because the value of a not pumped coin will remain at the same level as the purchase price and so do not produce an economic loss.
Instead, in our case, we want to provide a reliable approach---with high precision and recall--- to help investors stay out of the market when a pump and dump scheme is in action or to analyze anomalies in historical data.

\section{Related works}

The pump and dump phenomenon is older than the cryptocurrency revolution. Therefore, a wide portion of the literature is about pump and dumps done in the traditional stock market. Allen et al. in~\cite{allen1992stock} identify three categories of market manipulation schemes: information-based, action-based, and trade-based. The pump and dump schemes are usually a combination of information-based and trade-based manipulation.
In 2004, Mei et al.~\cite{mei2004behavior} show that it is possible to carry out pump and dump schemes just exploiting the behavioral biases of the investors. They test their theory on the pump and dump cases prosecuted by the SEC from 1980 to 2002, which confirm their hypothesis.

Several case studies highlighted that emerging markets were prone to pump and dump schemes.
%OLD Khwaja et al. in~\cite{khwaja2005unchecked} show that the Pakistani low regulation and the weak enforcement of the law on the national stock exchange allowed brokers to manipulate the market. They show evidence that the brokers of the Karachi Stock Exchange act for their own interests instead of working as intermediaries. They artificially raise the price of stocks trading among themselves. Once the prices rise they attract external traders and sell their stocks, leaving the traders with an asset that is work significantly less than what they paid for it.
Khwaja et al. in~\cite{khwaja2005unchecked} show that the Pakistani weak regulation of the national stock exchange allowed brokers of the Karachi Stock Exchange to arrange successful pump and dump schemes.
%OLD Jiang et al.~\cite{jiang2005market} investigate on the stock pools scheme of the '20s, which contributed to the development of the current anti-manipulation rules in the U.S.. The stock pools phenomenon is very similar to the pump and dump scheme. The stock pools are groups of traders that delegate to a single manager or an authority to trade stocks on their behalf and then they share the profits. Since a pool can move a large amount of money, they can increase the volume of trades and attract outsiders to the market. People start to buy thinking that something big is going on and prices rise. Then, when the stock pool exits the market, the price quickly drops. Jiang et al.\ assemble a new dataset with the daily trading volume from the New York Stock Exchange between 1927 and 1929, and they compare these data with the data of successful market manipulations. They find that the stock pools trade are more similar to informed trading than to a market manipulation scheme.
%such as the Securities Act of 1933, the Securities Exchange Act of 1934, and the creation of the SEC in 1934.
Jiang et al.~\cite{jiang2005market} investigate on the stock pools scheme of the '20s using daily trading volume from the New York Stock Exchange between 1927 and 1929. The stock pools are groups of traders that delegate to a single manager to trade stocks on their behalf. Since a pool can move a large amount of money, they can increase the volume of trades and attract outsiders to the market. When the stock pool exits the market, the price quickly drops. 
%OLD As reported by the University of Innsbruck in~\cite{frieder2007spam} the internet boom in the last years of the previous century and the early years of 2000 led to the birth of a new email based pump and dump scheme. In this new kind of fraud, the manipulators secure their position on the market and then send millions of e-mails claiming to have private information about strong increases in the prices of determined stocks. This causes an increase in the price and volume of the target stocks, followed by a drop as soon as the spam campaign ends. Of course, the drop is due to the manipulators who sell when the stocks prices reach their peak,having the losses to the standard investors.
 As reported by the University of Innsbruck in~\cite{frieder2007spam}, the Internet boom in the early years of 2000 led to the birth of a new email-based pump and dump scheme. In this new kind of fraud, the manipulators secure their position on the market and then send millions of e-mails claiming to have private information about substantial increases in the prices of determined stocks. This causes an increase in the price and volume of the target stocks, followed by a drop as soon as the spam campaign ends.
%OLD A successive analysis in 2013 by Siering in~\cite{siering2013all} shows that despite the authorities have taken several countermeasures against fraudulent stock recommendations, the problem is still far from being solved. Email based pump and dump campaigns are still successful. 
A later analysis in 2013 by Siering in~\cite{siering2013all} shows that despite the authorities have taken several countermeasures against fraudulent stock recommendations, email-based pump and dump campaigns are still successful. 

Some works about frauds in the cryptocurrency market have been published in the past few years.
%OLD The work of Gandal et al.~\cite{gandal2018price} show evidence that the first price spike to 1000 USD of the Bitcoin, may have been driven by a market manipulation. Using the well-known dataset of the Mt.Gox exchange they analyzed 18 millions Bitcoin transactions linked to user accounts. The most suspicious trading activities where carried out by two actors, named 'Willy bot' and 'Markus bot'. The purpose of these actors was to buy Bitcoin and so artificially increase the price and the daily volume. The two bots acquired a total of $600,000$ Bitcoins by November $2014$, moving on average respectively the $14$\% and $21$\% of the daily volume of Bitcoin trades. 
The work of Gandal et al.~\cite{gandal2018price} show evidence that the first price spike to 1000 USD of the Bitcoin may have been driven by market manipulation. Using the well-known dataset of the Mt.Gox exchange, they found suspicious trading activities carried out by two actors, named 'Willy bot' and 'Markus bot'. The purpose of these actors was to buy Bitcoin to increase the price and the daily volume artificially.
%OLD Krafft et al.~\cite{krafft2018experimental} investigate on the behavioral patterns of the users on the Cryptsy exchange market. In their work they show that even very small volumes of buy trades can influence the market. They use bots to buy a small amount of random currencies, and then observe the behavior of other users. They notice that the currencies being bought attract more traders than the usual. In particular, they conclude that traders tend to buy currencies with recent activities.
Krafft et al.~\cite{krafft2018experimental} investigate on the behavioral patterns of the users on the Cryptsy exchange market. In their work, they show that even tiny volumes of buy trades can influence the market. They use bots to buy a small amount of random currencies and conclude that traders tend to buy currencies with recent activities.
Li et al.~\cite{li2018cryptocurrency} conduct an empirical investigation on trading data obtained from the pump and dumps from Binance, Bittrex, and Yobit, focusing on the economic point of view. They show that pump and dumps lead to short term increase in prices, volume, and volatility followed by a reversal of the trend after some minutes. Moreover, they show that the gain of the investors depends critically on the time they obtain the signal, and for this reason, outside investors are systematically disadvantaged.
%Finally, an analysis of the Bittrex ban on pump and dumps, shows that they are detrimental to the health of the cryptocurrencies market.
%OLD The works of Xu et al.~\cite{xu2018anatomy} and work focuses on the difficult task of predicting pump and dumps, using one-hour intervals data from Cryptopia and Yobit. Finally, they report that their model in a real environment achieves $83.3\%$ of precision and recall of $18.5\%$.
%OLD Kamps et al.~\cite{kamps2018moon} showed a first approach to detect pump and dumps events through an adaptive threshold. They propose a set of features and build a detection algorithm that works on one-hour intervals. In their work, they raise the issue that does not exist a dataset of confirmed pump and dumps scheme, so they can not fully validate their results.
The work of Kamps et al.~\cite{kamps2018moon} shows a first attempt to detect pump and dumps using an adaptive threshold. They highlight the issue that a reliable dataset of the confirmed pump and dumps scheme does not exist, so they can not fully validate their results. A crucial contribution of our work is to release such a dataset. Xu et al.~\cite{xu2018anatomy} focuses on the difficult task of predicting pump and dumps, using one-hour intervals data from Cryptopia and Yobit, also showing an approach to exploit prediction to invest in alt-coins. Since both works have some goals in common with ours, we conducted a thorough analysis of their results on subsection~\ref{subsec:comparision}.
%As a future work they suggest to create a reliable confirmed pump and dump dataset to allow the validation of detection algorithms and the usage of supervised machine learning techniques.

\section{Discussion}

\textbf{Is it possible for pump and dump groups to avoid the detection?} We based our features on the idea of detecting an anomalous change of some market parameters and at the same time to be robust against the natural oscillations of the volatile cryptocurrencies market. So, if the admins of groups or some vip members start to buy the currency gradually in a way that the rise of this parameter is smooth, and the users are few, our classifier could be not able to detect the pump and dump. In this case, our classifier cannot detect $4$ of the pump and dumps in our dataset. These events were all carried out by one group, and all of them record a consistent pre-pump phase in the hours before the pump starts.
Fortunately, this technique cannot be applied frequently because a smart observer knowing the time of the next pump and dump can exploit this pattern to unveil in advance the targeted cryptocurrency. Moreover, after the pump and dump, this operation will be noticed by most users that could lose the trust of the admins and leave the group.\\
%This strategy guarantee to us an extremely high precision to detect pump and dump scheme but at the same time we can not  
\textbf{Can pump and dump groups manipulate Bitcoin or major cryptocurrencies?} To answer this question, we make a straightforward simulation. Let us take the buy volume on the first $10$ minutes of the biggest pump and dump we monitored that is of $31$ BTC on the SingularDTV (SNGLS). Now, we take a snapshot of the exchange order book for the trading pair BTC/USD (data of April 12, 2019), and assume that the market is frozen and only the members of the pump and dump group can make actions, this is the best case for them to make the price raise and get the attention of the outside investors. We find that the amount of money at their disposal can rise the value of BTC of less than $5$ USD. Now, we analyze the price oscillation of the Bitcoin in the hour before the snapshot on time ranges of $10$ minutes, and we find that the minimum oscillation was of $10$ USD while the maximum was more than $50$ USD. If we made another similar simulation on the Ripple (XRP), we get that the XRP value increases less than $0.0005$ USD, while the minimum and maximum oscillation is of $0.001$ and $0.002$. As we can see, only the volume of the orders in the orders book is enough to smooth the firepower of the pump and dump group, and the higher price peak they can achieve, in this extremely optimistic situation, is not enough to drum up the interests of outside investors.\\
\textbf{Is it possible for the exchange markets to stop pump and dump schemes?} In this work we show that it is possible to detect a pump and dump scheme as soon as it starts. We also believe that an exchange can detect better than us when a fraudulent scheme like this one is in action. In fact, the data owned by the exchange is more fine-grained than what we could obtain: It has full knowledge on the kind of operations performed, their amount, and exactly who performed them during the pump and dumps. Moreover, we notice that small policy enforcements against the pump and dumps can reduce drastically the amount of these market manipulations. As an example, BitTrex exchange announced~\cite{bittrex} that it actively discourages any type of market manipulation and will begin to punish the participants. Since then, the amount of pump and dumps in the exchange drastically decreased. We counted, before the statement, more than 50 pump and dumps in the 5 months from July to the end of November 2017, and only 31 in almost one year after the statement. Another possible counter-measure could be to stop transactions on a security when it gains or loses more than some threshold, or give special protection to cryptocurrencies with extremely low market capitalization and trading volumes.

\section{Conclusion}
In this work, we performed an in-depth analysis of the pump and dump ecosystem. %We defined a taxonomy of all the stakeholders involved in the market manipulation scheme, and 
We have studied the relationship that exists between the groups, the exchange, and the target cryptocurrencies. We also presented two case studies and a tool that can detect in real time a pump and dump in action. Moreover, we have identified a peculiar kind of orders that are particularly effective for the detection. %As a final contribution, we release to the community a dataset of confirmed pump and dumps.
We think that this work helps to understand a complex phenomenon, improve the awareness of the investors interested in the cryptocurrency market, and can help the authorities regulate this particular market in the future .

\section*{Acknowledgment}
This work was supported in part by the MIUR under grant ``Dipartimenti di eccellenza 2018-2022" of the Department of Computer Science of Sapienza University.
\bibliographystyle{./bibliography/IEEEtran}
\bibliography{./biblio}

\end{document}